\documentclass[conference]{IEEEtran}
\IEEEoverridecommandlockouts

\usepackage{cite}
\usepackage{amsmath,amssymb,amsfonts}
\usepackage{algorithmic}
\usepackage{graphicx}
\usepackage{textcomp}
\usepackage{xcolor}

\usepackage{verbatim} % weiping
\usepackage{threeparttable} % weiping
\usepackage{bm} % weiping
\usepackage{caption} % weiping
\usepackage{algorithmic} %weiping
\usepackage{algorithm} %weiping
\usepackage{multirow} %weiping
\usepackage{amsmath} %weiping
\usepackage{cite} %weiping

\def\BibTeX{{\rm B\kern-.05em{\sc i\kern-.025em b}\kern-.08em
    T\kern-.1667em\lower.7ex\hbox{E}\kern-.125emX}}

\usepackage{eso-pic}
\AddToShipoutPictureBG{%
  \AtPageUpperLeft{%
    \put(20,-20){\textbf{Accepted version for IEEE ICPADS-2025. \textcopyright~2025 Copyright held by the authors. Publication rights licensed to IEEE.}}%
  }%
}

\begin{document}

\title{A Bit Level Weight Reordering Strategy Based on Column Similarity to Explore Weight Sparsity in RRAM-based NN Accelerator\\
%{\footnotesize \textsuperscript{*}Note: Sub-titles are not captured for https://ieeexplore.ieee.org  and \%should not be used}
\thanks{DOI: https://doi.org/XXXXXXXX}
}

\author{Weiping Yang, Shilin Zhou, Hui Xu, Yujiao Nie, Qimin Zhou, Zhiwei Li, Changlin Chen\(^*\)\\\textit{College of Electronic Science and Technology, National University of Defense Technology, Changsha, China}}

\vspace{-0.5cm}
\maketitle

\begin{abstract}
Compute-in-Memory (CIM) and weight sparsity are two effective techniques to reduce data movement during Neural Network (NN) inference. However, they can hardly be employed in the same accelerator simultaneously because CIM requires structural compute patterns which are disrupted in sparse NNs. In this paper, we partially solve this issue by proposing a bit level weight reordering strategy which can realize compact mapping of sparse NN weight matrices onto Resistive Random Access Memory (RRAM) based NN Accelerators (RRAM-Acc). In specific, when weights are mapped to RRAM crossbars in a binary complement manner, we can observe that, which can also be mathematically proven, bit-level sparsity and similarity commonly exist in the crossbars. The bit reordering method treats bit sparsity as a special case of bit similarity, reserve only one column in a pair of columns that have identical bit values, and then map the compressed weight matrices into Operation Units (OU). The performance of our design is evaluated with typical NNs. Simulation results show a 61.24\% average performance improvement and 1.51×–2.52× energy savings under different sparsity ratios, with only slight overhead compared to the state-of-the-art design.
%Simulation results show an average performance improvement of 61.24\% and energy savings ranging from 1.51× to 2.52× under different sparsity ratios with slight overhead compared to the state-of-the-art design.
\end{abstract}

\begin{IEEEkeywords}
RRAM-based accelerator, weight reordering, bit level sparsity, bit level similarity.
\end{IEEEkeywords}

\vspace{-0.3cm}
\section{Introduction}
\vspace{-0.2cm}
Compute-in-Memory (CIM)\cite{cim23} architectures and weight sparsity\cite{sparsityNeuroC24,tcasi24} are two powerful techniques for reducing data movement and improving energy efficiency in Neural Network (NN) inference. CIM reduces data movement by storing data in non-volatile devices such as Resistive Random Access Memory (RRAM)\cite{rram21yu} , and weight sparsity techniques achieve NN compression by setting some unnecessary data to zero values. However, these two techniques are difficult to be utilized simultaneously, as CIM requires structured computational patterns that are disrupted by the irregular weight distribution of sparse NNs.

\begin{figure}[H]
\centering
\includegraphics[height=4.7cm]{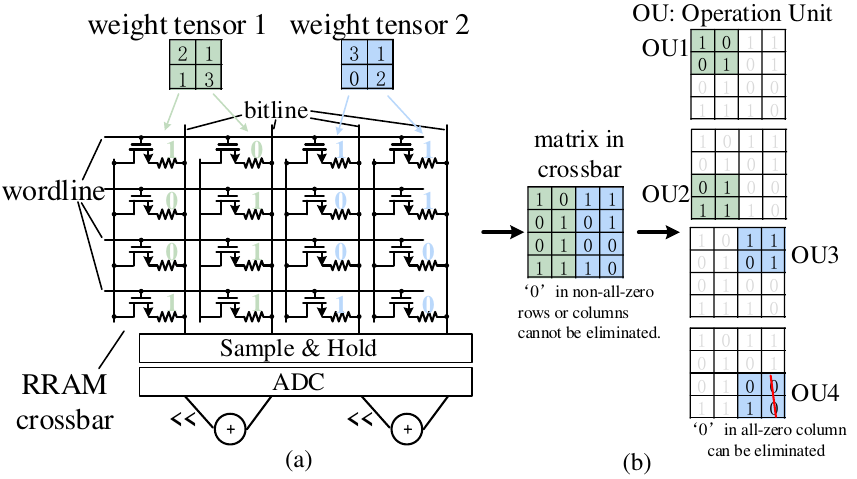}
\vspace{-0.3cm}
\caption{Weight mapping in RRAM crossbar and illustration of Operation Unit.}
\label{crossbarintro}
\end{figure}

The structured computational patterns in CIM architectures are based on two rules: each row in the CIM crossbars shares the same input, and each column generates one same output. Therefore, a single zero in a row or column cannot be removed unless the entire row or column contains only zeros.  A simplified example of the RRAM crossbar and weight mapping is illustrated in Fig. \ref{crossbarintro} (a), where two weight tensors are flattened into four column vectors in the crossbar, with each weight value represented by two RRAM cells in the same row. 
Although almost half of the cells store zeros, none of the rows or columns contain only zeros and thus no computation can be skipped, resulting in the waste of storage resources and power consumption. 
\vspace{-0.6cm}
\vspace{-0.6cm}

Existing weight sparsity or pruning techniques \cite{tpami24} can be broadly categorized into two types, coarse-grained pruning and fine-grained pruning.
Coarse-grained pruning, or structural pruning, is relatively hardware friendly since the zero values are continuous in one dimension of a pruned weight tensor, but the sparsity ratio is limited due to the trade-off between pruning rate and accuracy, leaving little room for acceleration. Fine-grained pruning has high sparsity ratios, yet the randomly distributed zero values are difficult to eliminate in the crossbar for efficient computation and storage. 

Existing works\cite{sre2019,recom2018,effective2022,Bflip2021,repim2021} that exploit NN sparsity in RRAM-based NN Accelerators (RRAM-Acc) mainly focus on eliminating zeros along either the row or column dimension in the crossbar at the Operation Unit (OU) level. An OU is a sub-region of the crossbar that can be activated during each clock cycle for practical and reliable computation. The small size of OU reduces the difficulty of exploiting the sparsity compared to the whole crossbar, as shown in Fig. \ref{crossbarintro}(b).

One idea behind these works is to prune or quantize the NNs with consideration of the crossbar structure\cite{recom2018,sre2019,Bflip2021}. The goal is to create as many all-zero rows or columns as possible. However, this process requires a considerable amount of time to modify and retrain the model.
Another idea is to innovate weight mapping strategies. This approach  aligns irregularly distributed zeros into the same rows or columns through reordering, without modifying the model\cite{effective2022, repim2021}. However, the benefits are limited, since only the bit sparsity in all-zero rows or columns can be exploited.

In this paper, we propose a bit level weight reordering strategy not only to utilize bit sparsity more extensively, but also to explore the bit similarity in non-all-zero columns.
According to our analysis and observation, there exists a high degree of similarity between the columns of the crossbar. With the proposed strategy, the repetitive all-zero columns at the OU level are eliminated, and only one column from each identical pair is retained.
Meanwhile, the two's complement format is employed to replace the conventional positive-negative weight splitting policy to store the weights so that the sparsity indexing overhead is reduced.
In addition, a weight splitting policy based on bit positions is also utilized to reduce the output indexing overhead.  A hardware architecture is also implemented to validate the effectiveness of our proposal.
The simulation results with five popular NNs demonstrate that our proposal achieves a 61.24\% performance improvement compared to the state-of-the-art design.
\vspace{-0.3cm}
\section{Background}
\vspace{-0.2cm}
Many endeavors have been performed to implement sparse NNs in RRAM-Accs, among which
Recom\cite{recom2018} and SNrram\cite{snrram2018} are the first two RRAM-Accs that support sparse processing. However, they only deal with the regular zeros produced by the coarse-grained pruning technique\cite{SSL2016} which has limited pruning rates. Meanwhile, the two designs have the entire 128\(\times\)128 crossbar involved in the computation within a single cycle, which is not a realistic situation due to the non-ideal effects of the RRAM crossbars.

Compared to Recom and SNrram, SRE\cite{sre2019} is a more practical design that proposes the OU-based computing paradigm. Calculations based on OU take into account the realities of the non-ideal effects of the RRAM crossbar and the hardware overhead of the Analog-to-Digital Converter (ADC), making them consistent with actual RRAM-Acc designs. Hence, most of the subsequent studies related to the RRAM-Acc hardware architecture employ an OU-based design, as is the case in this paper. However, SRE simply compresses the OU rows to bring very limited performance improvement, especially in the case of unstructured sparsity where the distribution of zero values is highly random. Nevertheless, SRE still achieves 15.8\% performance improvement compared to the over-idealized dense accelerator ISAAC\cite{isaac2016}.

Since the size of OU is smaller than the whole crossbar, it makes the generation of more all-zero rows or columns in OU level more easier even without modification to the models. Compared to SRE which only adopts row compression in OUs without optimization for weight mapping, RePIM\cite{repim2021} proposes a layer-wise weight exchange method to skip zero weights, and utilizes the repetitiveness of inputs and weights to remove redundant computations. RePIM is the state-of-the-art design which achieves 28\% performance speedup over SRE when reordering rows to achieve all-zero OU-column compression. However, RePIM only aims to produce all-zero OU columns. Therefore, there is still room for improvement in the performance gains. 

In common with RePIM, Hoon et al.\cite{effective2022} proposes a filter reordering scheme at the OU level that can effectively compress OU rows, but it requires NNs with very high sparsity ratios.  Compared to SRE, Hoon et al. achieves about 20\% performance improvement. The core idea behind RePIM and Hoon et al. is to aggregate all-zero OU rows or columns, but they fail to notice that some computations involving non-all-zero columns may be eliminated as well.

Apart from the above works, there are a number of other approaches that adopt training strategies to make the weights in the crossbar as repetitive as possible to eliminate redundant computations in RRAM-Accs\cite{Bflip2021}\cite{pattpim2020}\cite{prap_pim}\cite{rramsharing2021}. Pattpim\cite{pattpim2020} proposes an approximate weight pattern transform algorithm to increase the repetition of non-zero weight patterns, thereby reducing computation.  
BFLIP \cite{Bflip2021} reduces storage overhead by flipping bits in crossbar rows/columns to align with a centroid bit matrix (storing only the centroid), but still requires post-flipping calibration to maintain accuracy. 
Reram-sharing\cite{rramsharing2021} employs an alternating direction method of multipliers (ADMM)-based training approach to generate more identical small-bit matrices, thereby reducing storage and computational overhead. 
Generally, these works require complex modifications to the model, have limited applicability, and are time-consuming. Although these works do not consider sparsity, they do provide an idea of how repetitiveness of the weights can be exploited to eliminate computation overhead.

Beyond exploiting bit-level sparsity in RRAM-Accs, several design methodologies have been proposed to leverage bit-level sparsity in digital CMOS-based accelerator architectures. BitCluster \cite{bitcluster22} enforces each weight to have an identical number of one bits, simplifying the synchronization control logic induced by the irregular zero bits distribution and requires retraining. Bitlet \cite{bitlet21, bitlet23t} interleaves weights in computation array so that bit level sparsity can be exploited in parallel rather than in a bit-serial manner, eliminating the need for synchronization control logic. BitWave \cite{bitwave24} introduces a bit-column-serial approach with structured bit-level sparsity to maximize the utilization of computation units. BBS \cite{bbs24} leverages bit sparsity symmetrically to prune either zero bits or one bits, guaranteeing a sparsity level of more than 50\%, which also require post-training. BitWave \cite{bitwave24} and BBS \cite{bbs24} both require post-training. In summary, except for Bitlet, these architectures combine both hardware and software designs and require model training, while Bitlet’s approach is difficult to apply to crossbar-based RRAM-Accs. 

Despite these efforts in previous works, none of those works has theoretically analyzed the bit level sparsity or bit level similarity in models. In addition, none of them has made optimization in the storage format of the crossbar. The conventional positive-negative-split weight-mapping scheme which consumes a lot of crossbar resources also has room for optimization. 
More crucially, an overlooked insight is that computations can be eliminated not just by reordering to create rows or columns with all zeros, but also by reordering to create identical columns.
Based on these considerations, we propose a bit level weight reordering strategy that supports different sparsity ratios, and it works well at relatively low sparsity ratios compared to previous work\cite{effective2022}.

\vspace{-0.2cm}
\section{Weight Reordering}\label{WR}
\vspace{-0.2cm}
Our method is proposed based on the following observation: when weights are presented in two's complement format in matrices, a large portion — around 50\% — of the bits is 0, and many columns have similar bit patterns. When weight sparsity is taken into consideration, the percentage of '0's is much higher. We can squeeze out the 0 bits and repetitive bit patterns by swapping the rows or columns, so that only the minimum necessary weight bits remain in an OU.

To fulfill this target, we require that the weights be stored in the RRAM crossbar in two's complement format way. Compared to the positive-negative-split scheme or the bias scheme\cite{prime2016, isaac2016}, which incurs high crossbar resource overhead, the two's complement format saves 50\% of crossbar resources without retraining the models compared to previous work\cite{forms2021}.  Only a shift-and-subtract operation needs to be added to deal with the calculation of the sign bit. 

In detail, a decimal value $x$ can be represented by a $B$-bit binary value $x_{B-1}, x_{B-2}, \dots, x_0$ in two's complement format as shown in equation \eqref{twoscomplement}. 
\vspace{-0.3cm}
\begin{equation}
\label{twoscomplement}
x = -x_{B-1} \times 2^{B-1} + \sum_{i=0}^{B-2} x_{i} \times 2^{i}
\end{equation}
Mathmatically, the multiplication result of two signed \(B\)-bit data, \(IN_{[B-1:0]}\) and \(W_{[B-1:0]}\) , are expressed in equation \eqref{two2}. It can be observed from equation \eqref{two2} that the shift-and-subtract operation is required only when the second and third terms on the right-hand side of equation~\eqref{two2} are performed in the crossbar.
\vspace{-0.3cm}
\begin{align}
\label{two2}
IN_{[B-1:0]} \times& W_{[B-1:0]} = IN_{[B-1]} \times W_{B-1} \times 2^{2(B-1)}   &+ \nonumber \\
& (-IN_{[B-1]} \times 2^{(B-1)} \times \sum_{i=0}^{B-2}W_{[i]} \times 2^i)  &+ \nonumber \\
& (-W_{[B-1]} \times 2^{(B-1)} \times \sum_{i=0}^{B-2}IN_{[i]} \times 2^i)  &+ \nonumber\\
& (\sum_{i=0}^{B-2}IN_{[i]} \times 2^i \times \sum_{i=0}^{B-2}W_{[i]} \times 2^i)
\end{align}

\begin{figure}[H]
\centering
\includegraphics[height=5cm]{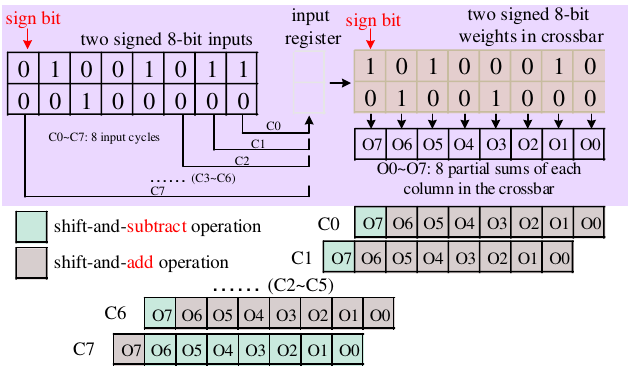}
\vspace{-0.3cm}
\caption{MAC operation in two's complement format.}
\label{shiftAS}
\end{figure}

Without loss of generality\cite{yang2020}, we set $B=8$ to explain our method. 
\(8\) RRAM cells in the same row across eight columns are utilized to represent one weight. 
The inputs are also quantized to \(8\) bits, which are fed to the crossbar bit by bit. 
Through shift-and-add/subtract operations, the partial sums are combined to form the final multiply-accumulate (MAC) results.  As shown in Fig. \ref{shiftAS}, MAC operations are perfomed between two inputs and two weights. Only O7 in the first seven cycles and O0--O6 in the last cycle undergo the shift-and-subtract operation, while the remaining partial sums undergo shift-and-add operations. Since subtraction is only required in the sign column of weights (with magnitude input bits), and for magnitude columns of weights (with sign input bit), the control logic remains simple, leading to minimal hardware overhead. In the subsequent analysis in this paper, the weights appear consistently in two's complement format.
\vspace{-0.3cm}

\vspace{-0.3cm}
\subsection{Bit Level Sparsity}
\vspace{-0.1cm}
In NNs, weight sparsity ratio is typically measured at the data level. NNs with a high sparsity ratio contain a large number of zero weights, making data-level sparsity exploitation potentially beneficial for hardware performance. However, in less sparse NNs, although zero-valued weights are rare, the number of zero bits remains significant and should not be underestimated. Assume that each value is represented using \(B\) bits in two's complement binary form. The binary representation of a zero value consists entirely of zeros, meaning all \(B\) bits are zero. For non-zero values, assuming the 0 bit and 1 bit are randomly distributed, each bit position has a 50\% probability of being 0 bit. Therefore, the proportion of 0 bits in \(B\)-bit \(N\) values is given by equation \eqref{bit0}, where \(p\) denotes the sparsity ratio of models.
\vspace{-0.2cm}
\begin{equation}
\label{bit0}
P_{0bit} = p + (1 - p) \times 0.5 = 0.5p + 0.5
\end{equation}
\vspace{-0.5cm}

\begin{figure}[H]
\centering
\includegraphics[height=5cm]{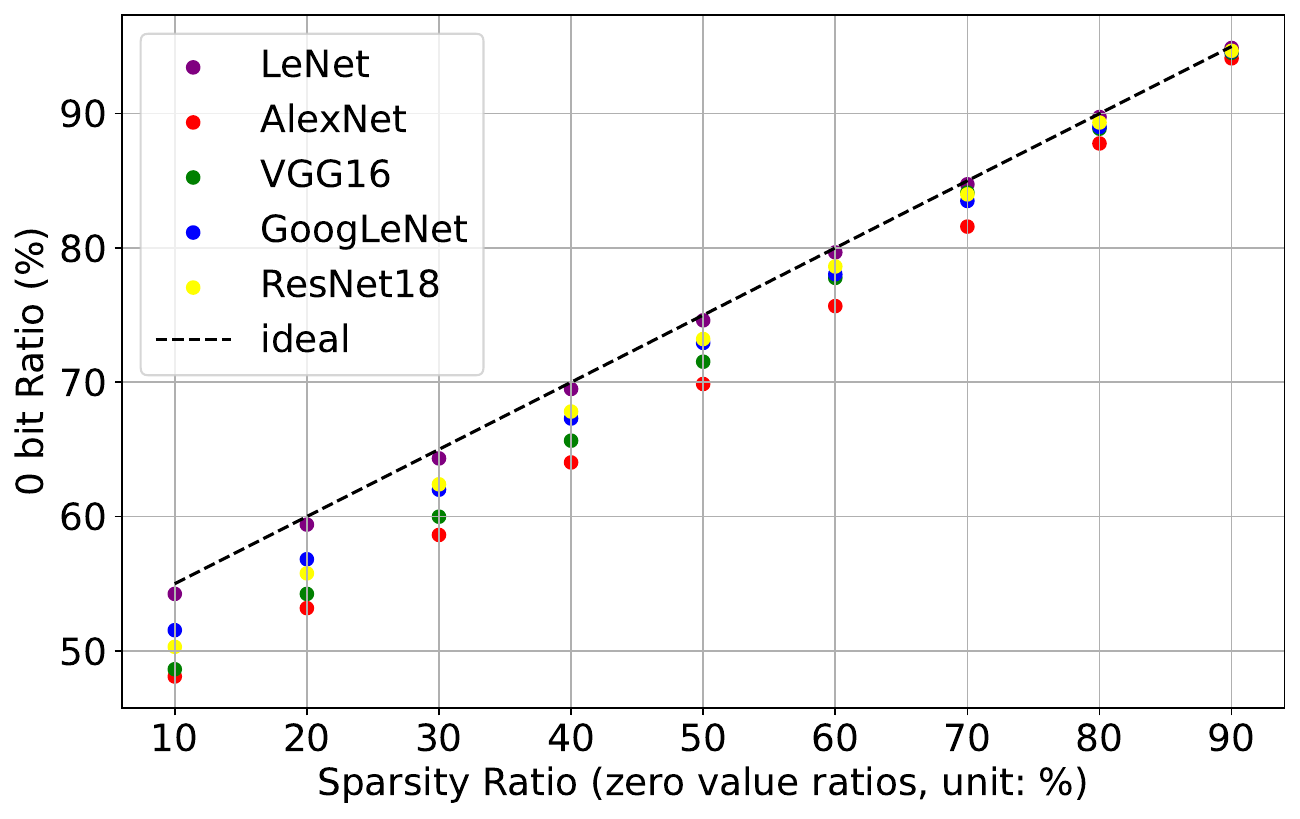}
\vspace{-0.1cm}
\caption{zero bit ratios in ideal situation and real models.}
\label{bit0ratio}
\end{figure}

To validate our analysis, experiments on several CNN models are conducted. The models' weights are first sparsified using the typical Lasso (L1) fine-grained pruning method, quantized to 8 bits, and then encoded with the binary complement method. The percentages of 0 bits in the weight tensors at different sparsity ratios are basically in line with the theoretical values in equation \eqref{bit0}, with only a slight decrease, as shown in Fig. \ref{bit0ratio}.
\vspace{-0.4cm}
\vspace{-0.5cm}
In general, sparsity at the bit level is impressive. Therefore, we could utilize the bit-level sparsity to remove the redundant calculations. However, even though the 0 bit ratios are at a high level in different sparsity ratios, only all-zero rows or columns could be utilized to eliminate the storage and calculations in the crossbar. For NNs with high sparsity ratios, the performance gain from utilizing 0 bit level sparsity may be effective, yet for NNs with relatively low sparsity ratios, the performance gain from utilizing bit-level sparsity is still a challenge. The bit level similarity is employed to mitigate this challenge. 
\vspace{-0.2cm}
\subsection{Bit Level Similarity}
\vspace{-0.5cm}
\begin{figure}[H]
\centering
\includegraphics[height=5cm]{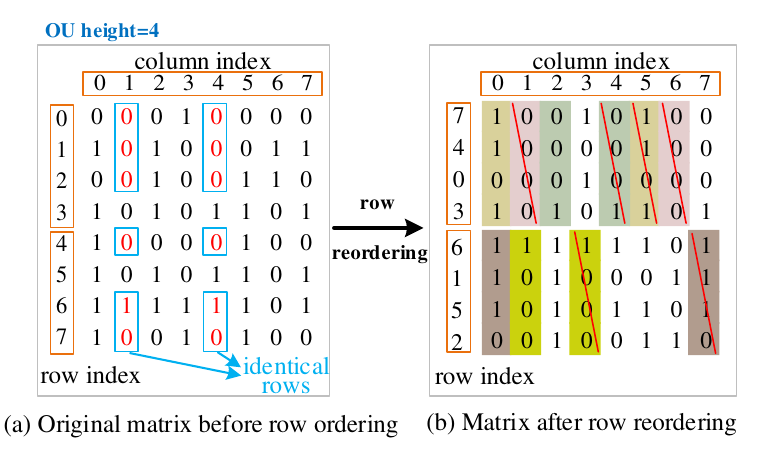}
\vspace{-0.3cm}
\caption{Example of row reordering to generate OU level identical column vectors.}
\label{colreorder}
\end{figure}
\vspace{-0.3cm}
An overlooked perspective is the existence of a large number of repetitive but discontinuous bits in the crossbar consisting of 0/1 bit, as is shown in red font in Fig. \ref{colreorder}(a). 

In Fig. \ref{colreorder}(a), there exist many identical rows in any two columns such as the Column 1 and Column 4. If these rows are reordered to form OUs, then more OU level identical column vectors can be obtained. In Fig. \ref{colreorder}(b), column vectors with the same base color are identical and only one column vector in each identical column vector pair needs to be stored.

It is worth noting that the identical column vector pairs include all-zero columns which could all be left unstored. Examples are Column 1 and Column 6 in Fig. \ref{colreorder}(b). This means that \textbf{reordering using bit level similarity has a higher efficiency than reordering based only on bit level sparsity}, and the former is more adaptable to the deployment needs of models with different sparsity ratios.

Generally, for the column vectors in the weight matrix composed of 0/1 bit, the probability of finding two identical column vectors is very low. 
However there is a high possibility that column vectors have the same bit value at the same row positions since each bit can only be 0 or 1. That is, those column vectors have the same value in some specific rows, but in general these rows are not contiguous. 
If the rows are reordered so that some consecutive rows have the same column vectors, then only at most one column vector needs to be computed and the results could be reused.

To further demonstrate the effectiveness of the above proposal, firstly a detailed mathematical analysis is given below prior to presenting the details of our algorithm. 
The problem could be abstracted as finding the number of identical positions with the same value in a set of \(n\) column vectors of length \(m\). The identical positions with the same value mean that the values are either all '0's or all '1's in the same row.

In a set of \(n\) column vectors denoted by $a^{(1)}$, $a^{(2)}$,..., $a^{(n)}$, since the value at each position in the vectors are independent and have equal probability of being '0' or '1', the probability that these vectors are all 0s and all 1s at position \(i\) is expressed by equation \eqref{eq_prob0}, where \(i\) denotes the row index.
\begin{equation}
P(a_i^{(1)} = a_i^{(2)} = \ldots = a_i^{(n)} = 0 \mbox{ or } 1) = 2 \times \frac{1}{2^n} = \frac{1}{2^{n-1}}\label{eq_prob0}
\end{equation}
Therefore, the probability of finding an identical row in \(n\) column vectors is described as $p=\frac{1}{2^{n-1}}$. Define \( X \) as the number of identical rows among \( n \) column vectors of length \(m\), \( X \) obeys a binomial distribution with parameters \( m \) and \( p = \frac{1}{2^{n-1}} \). The formula is shown in equation \eqref{binomial01},  where \(k = 0, 1, 2, \ldots, m\), \(C_m^k = \frac{m!}{k!(m-k)!}\).
\begin{equation}
\label{binomial01}
P(X = k) = C_m^k (\frac{1}{2^{n-1}})^k ( 1 - \frac{1}{2^{n-1}})^{m - k}
\vspace{-0.1cm}
\end{equation}
Then the probability that \(n\) column vectors of length \(m\) have \textbf{at least} \(k\) identical rows could be computed using the cumulative distribution function of the binomial distribution, which results in equation \eqref{sum_k}.
\vspace{-0.3cm}
\begin{equation}\label{sum_k}
P(X \geq k) = 1 - \sum_{i=0}^{k-1} C_m^i (\frac{1}{2^{n-1}})^i ( 1 - \frac{1}{2^{n-1}})^{m-i}
\vspace{-0.1cm}
\end{equation}
Equation \eqref{sum_k} contains three parameters: \(k\), \(m\) and \(n\). With \(n\) taken as 2 and \(k\) taken as \(m/2\), equation  \eqref{sum_k} reduces to equation \eqref{sum_m2}
\vspace{-0.4cm}
\begin{equation}\label{sum_m2}
P(X\geq\frac{m}{2})=1 - \sum_{i=0}^{\frac{m}{2}-1} C_m^i \frac{1}{2^m}
\vspace{-0.2cm}
\end{equation}
Equation \eqref{sum_m2} implies that for two column vectors (\(n=2\)), the probability that at least half of the rows are identical is greater than \(50\%\). These large number of identical column rows could be aggregated to form identical column vectors by row reordering. Therefore, in this paper, we choose \(n=2\) to ensure a higher probability of finding identical rows.

\begin{figure}[H]
\centering
\includegraphics[width=9cm]{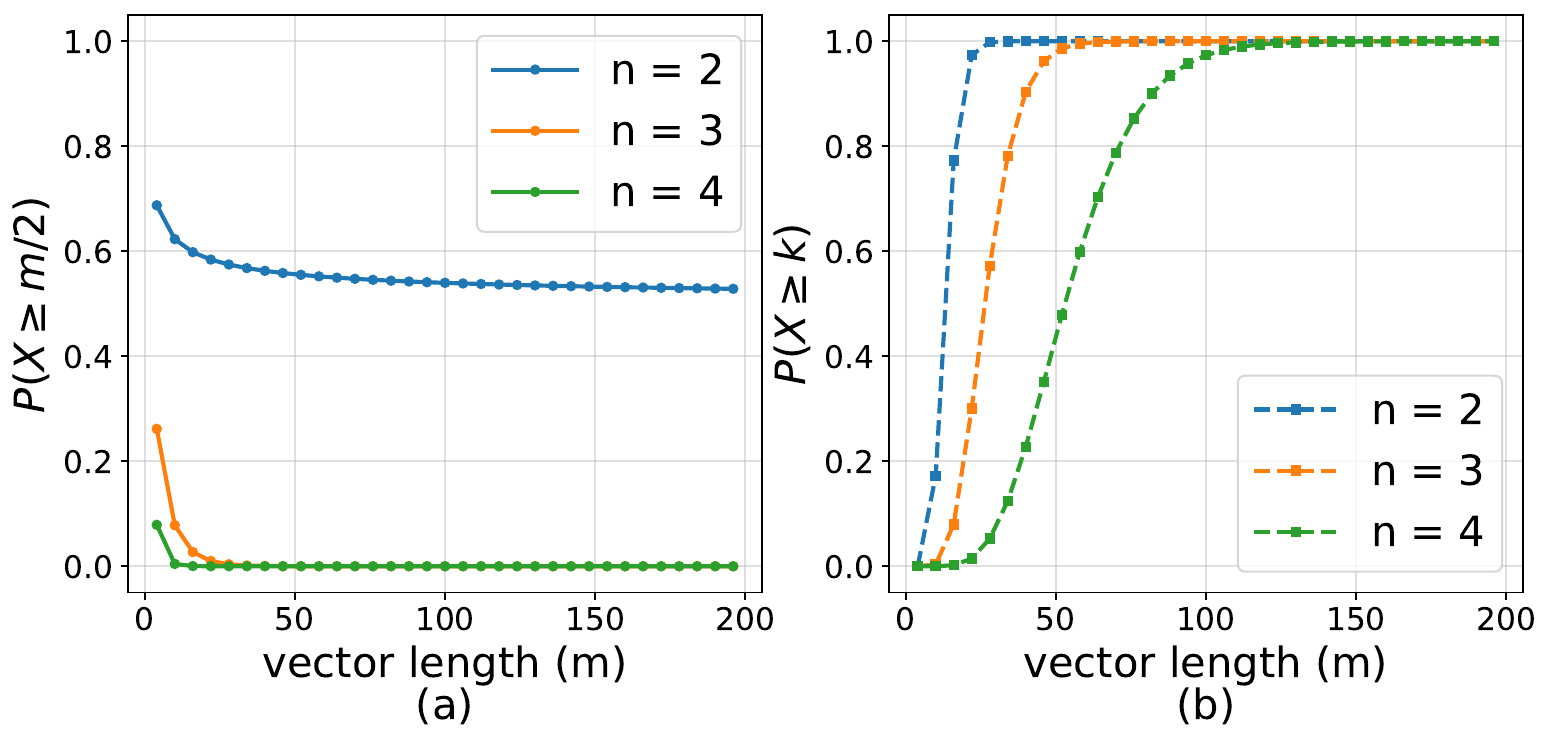}
\caption{(a) Probability of at least half identical rows.
    (b) Probability with fixed identical rows \(k\)=7.}
\label{mnk}
\end{figure}

In detail, according to equation \eqref{sum_k}, a more comprehensive analysis of extending bit level similarity to the grouping of three or more columns is illustrated in Fig.~\ref{mnk}. As shown in Fig.~\ref{mnk}(a), when \(n > 2\), as in the case of \(n = 3\), the probability of having at least half of the rows identical drops to no more than 30\% and continues to decrease as \(n\) increases. Moreover, as the vector length \(m\) becomes larger, this probability approaches zero. For \(n > 4\), similar trends are observed. In Fig.~\ref{mnk}(b), for a fixed number of identical rows \(k = 7\), a larger \(m\) is required to achieve the same probability with increasing \(n\), indicating that finding identical rows becomes increasingly unlikely for larger \(n\).
\vspace{-0.2cm}

\subsection{Proposed Reordering Algorithm}
Based on the aforementioned analysis, we propose a row reordering algorithm to improve the column vector repetition rate at the OU level, aiming to boost RRAM crossbar resource utilization and computational energy efficiency by eliminating redundant computations. 

We first utilize Hamming Distance to calculate the similarity between any two equal-length column vectors based on the number of identical rows they have.  The concrete formula is shown in  equation \eqref{sHD}. \(Va\) and \(Vb\) denote two equal-length column vectors. \(m\) denotes the length of vectors. 
\(sHD\) denotes the number of different rows of the two column vectors, where \(s\) is derived from the initial letter of \underline{s}imilarity.
Smaller values of $sHD$ imply that the vectors are more similar to each other and have more identical rows.
\vspace{-0.1cm}
\begin{equation}
\label{sHD}
sHD(Va, Vb) = \sum_{i=0}^{m-1} XOR (Va_i, Vb_i)
\vspace{-0.3cm}
\end{equation}

Based on \(sHD\), we propose our bit level weight reordering algorithm which is shown in Algorithm \ref{reorder_same} and Algorithm  \ref{reorder_same2}. The basic idea in the two algorithms is to find column vector pairs with the most number of identical rows step by step, and adjust these identical rows from discontinuous to continuous.

Algorithm \ref{reorder_same} processes a binary matrix \(M_{m \times n}\). It works by calculating the \(sHD\) between all column vectors. First, it finds column vector pairs with the smallest \(sHD\), the corresponding column indices are recorded first. Next, it finds column vector pairs with the second smallest \(sHD\). This process continues until all columns are paired. The output is a dictionary \(D\). Each key in \(D\) represents the column indices of column vector pairs. The corresponding value stores identical row indices (\(rowid\)) and the number of identical rows (\(numrows\)) in the column vector pair. 

\vspace{-0.3cm}
\begin{algorithm}[H]
    \caption{\(column\_pair\): Get identical column vector pairs}
	\label{reorder_same}
    \renewcommand{\algorithmicrequire}{\textbf{Input:}}
    \renewcommand{\algorithmicensure}{\textbf{Output:}}
    \begin{algorithmic}[1]
   	\REQUIRE Matrix \(M_{m\times n}\) %with \(m\) rows and \(n\) columns composed by 0/1 bit
   	, column/row indices set \(S_c\)/\(S_r\)

   	\ENSURE Dictionary \(D\)
   	
   	\WHILE{\(length(S_c) \geq 2\) }
   	\STATE $minshd \gets \infty$
	 	%\STATE \( minshd \leftarrow min(sHD(M[:,i], M[:,j]))\). \(i, j \in S_c\text{,} i \neq j\)
	 	
	 	\FOR{$i \in S_c$}
	 	    \FOR{$j \in S_c \text{ and } j > i$} 
	 	        \STATE $current\_sHD \gets sHD(M[:,i], M[:,j])$
	 	        \IF{$current\_shd < minshd$}
	 	            \STATE $minshd \gets current\_shd$
	 	            \STATE $(i^*, j^*) \gets (i, j)$
	 	        \ENDIF
	 	    \ENDFOR
	 	\ENDFOR
        	
        \STATE  \(numrows \leftarrow m - minshd\)
    	 \STATE \(mask \leftarrow\)  \(bitwiseXOR (M[:,i^*], M[:,j^*])\)

        	\STATE \(rowid \leftarrow S_r[ mask\mbox{==}0]\), where '0' in \(mask\) indicates the rows are identical
        	\STATE \(S_c \leftarrow S_c - \{i^*, j^*\}\)
        	\STATE \(D \xleftarrow{\mbox{append}} \) \{\((i^*, j^*):[rowid, numrows]\)\}
   	\ENDWHILE
   	\RETURN \(D\)
	\end{algorithmic}
\end{algorithm}

Algorithm \ref{reorder_same} is utilized as a function called \(column\_pair\) in the Algorithm \ref{reorder_same2}.
\(column\_pair\) groups the column vectors of the input matrix into column vector pairs according to the value of \(sHD\). Then the column vector pairs are utilized as the initial selected pairs iteratively in Algorithm \ref{reorder_same2} to search for the identical column vector pairs with the same input at the OU level. In Algorithm 2, \(numrows\) will gradually decrease from a larger value to the \(OU_{height}\), and the search stops when \(numrows\) is smaller than the \(OU_{height}\).

\begin{algorithm}[H]
 \caption{Reordering towards bit column repetitiveness}
 \label{reorder_same2}
     \renewcommand{\algorithmicrequire}{\textbf{Input:}}
     \renewcommand{\algorithmicensure}{\textbf{Output:}}
     \begin{algorithmic}[1]
         \REQUIRE Matrix $M_{m \times n}$, OU shape $h \times w$.
         \ENSURE Row indices List $L_{R}$ after reordering, List $L_{C}$ to store the indices pairs of identical column vectors in OUs.
         \STATE \textbf{Initialize} \(S_r\leftarrow \{0, 1, \ldots, m-1\}\), \(S_c\leftarrow  \{0, 1, \ldots, n-1\}\)
		 \STATE \(MS \leftarrow M_{m \times n}\);
         \(MS_r \leftarrow S_r\), \(MS_c \leftarrow S_c\)
         
         \WHILE{\(length(S_r) \geq h\)}

         \STATE \(D' \leftarrow column\_pair(MS, MS_c, MS_r)\)\\ 
         
         \FOR{each \{\((i, j):[rowid, numrows]\)\} $\in D'$}
         	\STATE \textbf{Initialize} an empty list \(L_{pair}\) to store the column indices of identical column vectors in one OU
         	
               	%\STATE Remove \(i\) and \(j\) from \(S_c\), Append \((i, j)\) to \(L_{pair}\)
               	\STATE \(S_c \leftarrow S_c - \{i, j\}\); \(L_{pair} \xleftarrow{\mbox{append}} (i, j)\)
               	
               	%\STATE Selected Matrix is $MS \leftarrow M[rowid,:]$, which has \(numrows\) rows and \(n\) columns; %List \(rowid\_ou\) to store the row indices after reordering.
               	\STATE $MS \leftarrow M[rowid,:]$
               	
               	\WHILE{$numrows \geq h$}
               	
 	             \STATE \( minshd \leftarrow min(sHD(C_a, C_b))\). \(a, b \in S_c, a \neq b\).\\ \(C_a\), \(C_b \in \mbox{column vectors of } MS\)
 	              \STATE \(mask \leftarrow bitwiseXOR(C_a, C_b)\)
 	             \STATE  \textbf{Update} \(numrows \leftarrow numrows - minshd\)
 	             %\STATE $\mathbf{numrows \leftarrow numrows - minshd}$

 	             \IF{\(numrows \geq h\)}
 	             \STATE \(rowid \leftarrow rowid[mask\mbox{==}0]\), where '0' in \(mask\) indicates the rows are identical
 	             \STATE \(MS \leftarrow M[rowid, :]\)
 	             \STATE \(S_c \leftarrow S_c - \{a, b\}\); \(L_{pair} \xleftarrow{\mbox{append}} (a, b)\)
 	             %\STATE Remove \(a\), \(b\) from \(S_c\). Append \((a, b)\) to \(L_{pair}\)
 	             \ELSE
 	             %\STATE There is no identical column vectors cater to $OU_{height}$, \(rowid \leftarrow rowid[:h]\) then save the result: \(D_{OU}[(i, j)] \leftarrow [rowid, L_{pair}]\)
 	             \STATE \(rowid \leftarrow rowid[:h]\)
 	             \STATE \(D_{OU}[(i, j)] \leftarrow [rowid, L_{pair}]\)
 	             \ENDIF
 	             
               	\ENDWHILE
         \ENDFOR
          \STATE \(L_C \xleftarrow{\mbox{append}} L_{pair}\) with \(max(length(L_{pair})\) in \(D_{OU}\); 
         \STATE \(L_R \xleftarrow{\mbox{append}} rowid\) with \(max(length(L_{pair})\) in \(D_{OU}\);
		\STATE  \(S_r \leftarrow S_r - rowid\)
		\STATE Update \(MS_r \leftarrow S_r\), \(MS_c \leftarrow S_c\)

      \ENDWHILE
      \RETURN \(L_R\), \(L_C\) 
         
     \end{algorithmic}
\end{algorithm}

\begin{figure}[H]
\centering
\includegraphics[width=8.8cm]{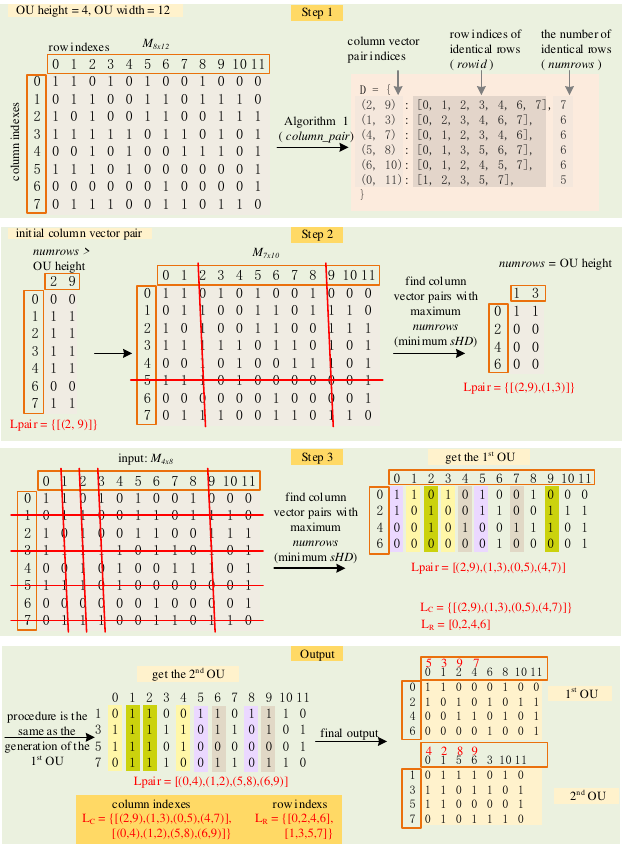}
\vspace{-0.7cm}
\caption{A typical example of the proposed bit level reordering strategy}
\label{walkthrough}
\end{figure}

A typical realization of the proposed algorithm is shown in Fig. \ref{walkthrough}. 
In Step 1 of Fig.\ref{walkthrough}, the column vector pairs indices are first identified in the initial matrix \(M_{8 \times 12}\) by Algorithm \ref{reorder_same}, followed by recording and ranking the row indices of identical rows. 
In Step 2, the column vector pair (2, 9) which has the most number of identical rows is chosen as the initial column pair, itself is removed from the \(M_{8 \times 12}\), and the identical row indices of column vector pair (2, 9) are utilized to select the rows to form the new matrix \(M_{7 \times 10}\). Then \(M_{7 \times 10}\) are utilized to find the new column vector pair, and column pair (1, 3) is selected. Since the number of identical rows are equal to the \(OU_{height}\), then the row indices of the first OU is determined. In Step 3, the column indices of the column vector pairs are recorded. Finally, other OUs are generated with the same procedure in Step1--Step3. The final column pairs indices and row indices are recorded in \(L_C\) and \(L_R\) respectively.
\vspace{-0.5cm}
\vspace{-0.5cm}
When applying the reordering algorithm to real models, attention needs to be paid to the versatile shape of the weights in a wide variety of model structures. The diverse architectures of neural networks need different numbers of crossbars. When the weight matrix is too large, the weight matrix needs to be split to improve the execution efficiency of the reordering algorithm and to fit the size of the crossbar, which also simplifies the indexing circuit.

\begin{figure}[H]
\centering
\includegraphics[height=6.5cm]{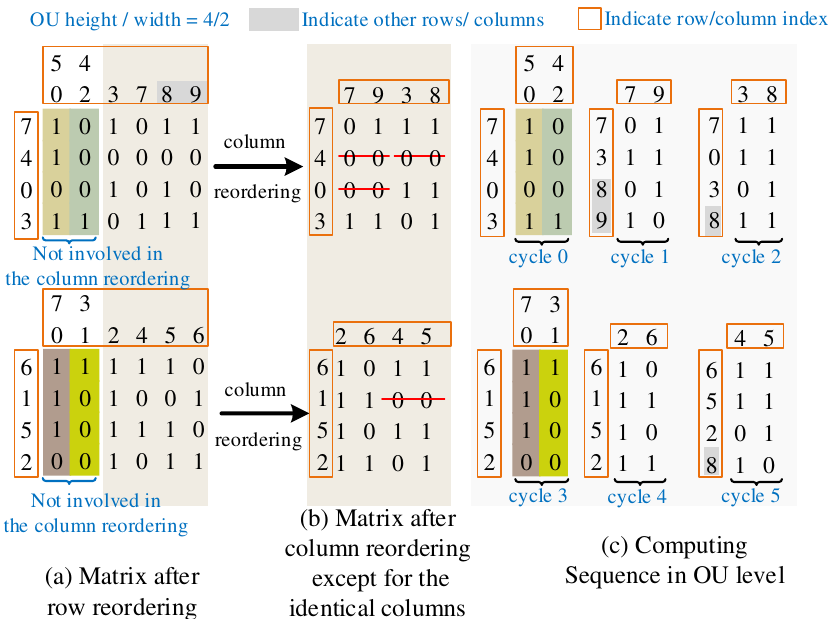}
\caption{Row compression procedure and computing sequence at the OU level.}
\label{rowreorder}
\end{figure}

A detailed example to compress the weight matrices with the proposed method is presented in Fig. \ref{rowreorder}. Note that the rows with all zeros are also compressed.
The reordering process begins with rearranging rows to create identical column pairs, followed by reordering columns to generate all-zero rows. When applying the reordering algorithm to gather all-zero rows by rearranging columns, identical column pairs remain unchanged and are excluded from the column reordering, as illustrated in Fig. \ref{rowreorder}.
The all-zero rows will be removed, and other not-all-zero rows are moved up to fill in the empty spaces after the elimination of all-zero rows. The row compression algorithm could be adopted from the scheme in RePIM\cite{repim2021}.

\vspace{-0.2cm}
\vspace{-0.5cm}

\section{Hardware Design \& Simulation Results}
\begin{table}[htbp]
\vspace{-0.2cm}
\caption{Hardware Configuration Details}
\begin{tabular}{|c|c|c|c|c|}
\hline
\multirow{2}{*}{\begin{tabular}[c]{@{}c@{}}\textbf{Works} \&\\ \textbf{Specs} \end{tabular}}
 & \multirow{2}{*}{\textbf{Ours}}
 & \multirow{2}{*}{\begin{tabular}[c]{@{}c@{}} \textbf{RePIM}\\ \textbf{\cite{repim2021}} \end{tabular}}
 & \multirow{2}{*}{\begin{tabular}[c]{@{}c@{}} \textbf{SRE}\\ \textbf{\cite{sre2019}} \end{tabular}}
 & \multirow{2}{*}{\begin{tabular}[c]{@{}c@{}} \textbf{Hoon}\\ \textbf{et al. \cite{effective2022}} \end{tabular}} \\
&  &    &   &  \\ \hline
\textbf{Bits-per-cell}  & \(1\) bit  & \(1\) bit  & \(2\)bit & \(2\)bit  \\ \hline
 
\multirow{2}{*}{\begin{tabular}[c]{@{}c@{}}\textbf{ADC}\\ \textbf{Resolution}\end{tabular}} & \multirow{2}{*}{\(3\) bit} & \multirow{2}{*}{\(4\) bit} & \multirow{2}{*}{\(6\) bit} & \multirow{2}{*}{\(6\) bit} \\
& & &&\\ \hline
 
\multirow{2}{*}{\begin{tabular}[c]{@{}c@{}} \textbf{Readout}\\ \textbf{Rows} \end{tabular}}
 & \multirow{2}{*}{\(\leq7\) } & \multirow{2}{*}{\(\leq 15^*\) } &  \multirow{2}{*}{\(\leq 21^*\) } & \multirow{2}{*}{\(\leq 21^*\) } \\
&&&&\\ \hline                                                                             
\textbf{OU Size}  & \(7 \times 8\)  & \(8 \times 8\)   & \(16 \times 16\)  & \(16 \times 16\)  \\ \hline
\textbf{Crossbar Size}&  \(128\times128\)  &   \(128 \times 128\) &  \(128 \times 128\)  & \(128 \times 128\)\\ \hline

\multirow{2}{*}{\begin{tabular}[c]{@{}c@{}}\textbf{Weight }\end{tabular}} & \multirow{2}{*}{\begin{tabular}[c]{@{}c@{}} 8-bit \\int \end{tabular}} & \multirow{2}{*}{\begin{tabular}[c]{@{}c@{}} 11-bit\\float \end{tabular}} &
\multirow{2}{*}{NaN} &
\multirow{2}{*}{\begin{tabular}[c]{@{}c@{}} 16-bit \\fixed point \end{tabular}} \\
 && && \\ \hline
  
\multirow{2}{*}{\begin{tabular}[c]{@{}c@{}}\textbf{Activation}\end{tabular}} & \multirow{2}{*}{\begin{tabular}[c]{@{}c@{}}8-bit \\int\end{tabular}} & \multirow{2}{*}{\begin{tabular}[c]{@{}c@{}} 11-bit \\float \end{tabular}} &
\multirow{2}{*}{NaN} &
\multirow{2}{*}{\begin{tabular}[c]{@{}c@{}} 16-bit \\fixed point \end{tabular}} \\
  && && \\ \hline
  
\multirow{2}{*}{\begin{tabular}[c]{@{}c@{}}\textbf{Simulator }\end{tabular}} & \multirow{2}{*}{Custom} & \multirow{2}{*}{\begin{tabular}[c]{@{}c@{}}NeuroSIM\\ \cite{neurosimv22021} \end{tabular}} & \multirow{2}{*}{Custom} & \multirow{2}{*}{Custom} \\
  && && \\ \hline
  
\hline \multicolumn{5}{c}{\textbf{Power Details in 1.2GHz 32nm process}}    \\ 
\hline \multicolumn{2}{|c|}{Components} & Metric & \multicolumn{2}{|c|}{Power}\\ 
\hline \multicolumn{2}{|c|}{One DAC} & 1bit & \multicolumn{2}{|c|}{0.049mW} \\ 
\hline \multicolumn{2}{|c|}{One ADC} & 3bit & \multicolumn{2}{|c|}{6.05mW} \\ 
\hline \multicolumn{2}{|c|}{A one-bit readout circuit} & 1bit & \multicolumn{2}{|c|}{0.2mW} \\ 
\hline \multicolumn{2}{|c|}{One Shift-and-Add circuit} & not care & \multicolumn{2}{|c|}{7.29mW} \\
\hline \multicolumn{2}{|c|}{Buffer in Computation Unit} & 128B & \multicolumn{2}{|c|}{4.2mW} \\
\hline
\end{tabular}
			\begin{tablenotes}
				\item[] $^*$ Works have underutilization of the ADC resolution since the readout rows are bigger than the $OU_{height}$
			\end{tablenotes}
\label{hardware_detail}
\end{table}
\vspace{-0.3cm}
A custom simulator realized in Python is designed to evaluate the effectiveness of our proposal. Five typical benchmarks, LeNet5-MNIST, AlexNet-ImageNet, VGG16-ImageNet, GoogleNet-ImageNet and ResNet18-ImageNet, are employed. 
Starting with a pretrained model, we apply the L1 unstructured pruning provided by PyTorch, which is a common approach used in many studies.
Then, the weights in the sparse model are quantized to signed 8-bit data \cite{quant8b21} using the Post-Training Quantization (PTQ) algorithm. Finally, the weights are represented in two's complement format. The hardware configurations are consistent with RePIM, with slight differences shown in Table \ref{hardware_detail}. The power data of the key modules in the lower part of Table \ref{hardware_detail} were obtained using Cadence Virtuoso and Synopsys Design Compiler in a 28nm process, and then the data were scaled to a 32nm process for ease of comparison with other works.

%In addition, it should be noted that this paper  does not consider hyperscale networks such as Transformers. This is because Transformers require dynamic data updates, whereas in RRAM-Accs, weights are usually stored statically because weight updates are both time- and energy-consuming\cite{tedwrite12}. Furthermore, most scenarios using RRAM-Accs involve power-limited edge computing devices.

%In addition, it should be noted that this paper though focuses on the how the method could adapt to dynamic or attention-based models, even if they are not the current focus.

In addition, it should be noted although hyperscale models such as Transformers are not considered in this work due to their dynamic data updates, our proposed approach could potentially be adapted to such architectures. For instance, hybrid memory schemes\cite{hybrid22TCAD,hyblearn24,sramRom25} combining RRAM with faster, volatile memory could enable partial or selective weight updates to support attention mechanisms. Exploring such adaptations would be an interesting direction for future work.

\subsection{Sensitivity Analysis}

In our reordering algorithm, the size of the OU directly affects the performance of the algorithm. Refer to equation \eqref{sum_k}, the parameter \(k\) actually is the $OU_{height}$.
The smaller the $OU_{height}$, the higher the probability of finding identical column vectors and the higher the compression ratio. The compression ratio is defined as the ratio of required computational crossbar quantities (CCQ) after reordering to the original required CCQ before reordering.

\vspace{-0.3cm}
\begin{figure}[H]
\centering
\includegraphics[height=6.5cm]{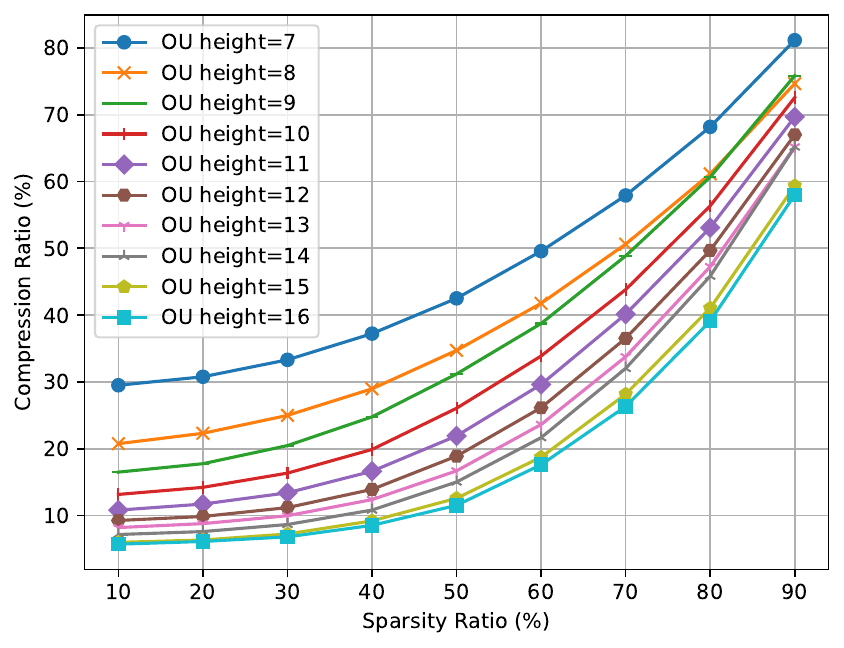}
\caption{Compression ratio of LeNet5 at different sparsity ratios under different $OU_{height}.$
}
\vspace{-0.2cm}
\label{lenet5ourange}
\end{figure}
\vspace{-0.3cm}

Fig. \ref{lenet5ourange} illustrates the crossbar resource compression ratio of the LeNet5 for different $OU_{height}$. It could be clearly seen that the compression ratio decreases as the $OU_{height}$ increases, which is also consistent with the previous analysis. The trend is also similar in other models and values of $OU_{height}$. 

While  $OU_{height}$ is also limited by the resolution of the ADC. A smaller $OU_{height}$ requires a lower ADC resolution, which results in lower hardware overhead. Even though smaller $OU_{height}$ means more computing cycles, it also implies shorter cycle period. Considering the state-of-the-art RRAM-based readout circuit\cite{tdc8b}, the resolution of ADC is set to 3 bit, which means that $OU_{height}$ should be set to 7.

As for the $OU_{width}$ which are constrained by the number of ADCs, the larger the $OU_{width}$, the more difficult it is to gather all-zero OU rows, and the less efficient the compression is. While at the same time, the input indexing overhead is reduced. For example, if $OU_{width} == Crossbar_{width}$, then OUs on the same row share the same inputs, and the inputs reordering overhead is reduced compared to the situation when OUs on the same row have different inputs. 
On the other hand, the smaller the $OU_{width}$, the more prone it is to have OU rows with all zeros, and therefore the more efficient the compression is. To ensure a fair comparison, the $OU_{width}$ in this paper is consistent with the most advanced design with details found in Table \ref{hardware_detail}, where $OU_{width}=8$.

\subsection{Indexing Overhead}\label{indexOverhead}

\vspace{-0.3cm}
\begin{figure}[H]
\centering
\includegraphics[width=9cm]{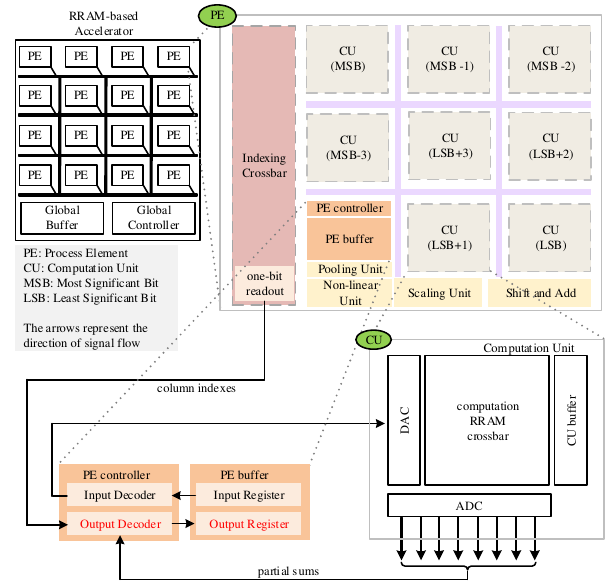}
\caption{Proposed hardware architecture and details of weight splitting \& output indexing.}
\label{control}
\end{figure}
\vspace{-0.3cm}
Since the reordering algorithm will disrupt the position of the weight bits, if the higher-order bits to lower-order bits of the weights are stored within the same crossbar, then additional indexing resources are needed to record the shift values. To simplify the output addressing, we propose a bit splitting policy.

Specifically, each layer's weight is decomposed into binary bits from Most Significant Bit (MSB) to Least Significant Bit (LSB). Unlike the traditional mapping where all bits of each weight are in the same crossbar,  the bits at different positions of a single weight are located in different crossbars in this paper. The advantage is that all the results of the computation in the same crossbar can be shifted by the same value, which reduces the overhead of the storage resources that are used to record the shifted values. As for the load imbalance issues which are commonly exist in sparse RRAM-Accs even without utilizing bit splitting policy, SPCIM\cite{spcim21} and Dyn-bitpool\cite{dyn25} have already proposed solutions which could be employed in this paper.
Moreover, because there is no need to record shift values, the indexing overhead in this paper is reduced by 10\%-31\%  compared to the state-of-the-art RePIM.
The realization of the weight splitting policy is shown in the Processing Element level (PE) of Fig. \ref{control}, where eight bits in one weight value are placed at eight different Computation Units (CU).

\begin{figure}[H]
\centering
\includegraphics[width=8cm]{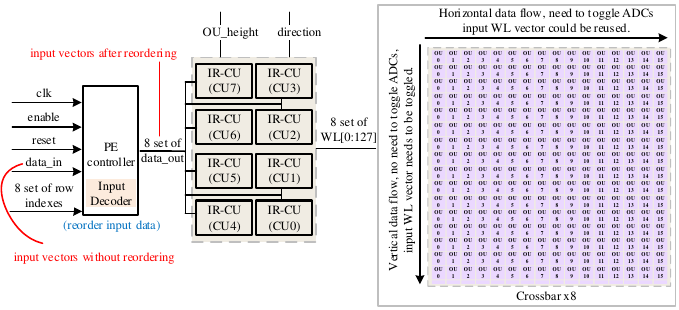}
\vspace{-0.2cm}
\caption{Input Routing Logic.}
\label{input_load1020}
\end{figure}

\textbf{Input Routing Logic:} The input indexing is handled by the Input Decoder in PE controller, as shown in Fig. \ref{control} and Fig.~\ref{input_load1020}. The Input Decoder module fetches input vectors from the PE buffer sequently and reorders the input vectors and distributes them to eight CUs. Our reordering logic follows the same approach as prior works requiring input indexing\cite{recom2018}\cite{effective2022}. The fundamental principles remain consistent with existing methods. 
\vspace{-0.4cm}
\vspace{-0.5cm}
However, it is notable that in our architecture, considering the value of $OU_{width}$, we design two computation directions: horizontal and vertical, to flexibly adapt to different OU configurations. Details are shown in Fig.~\ref{input_load1020}. In our design, the calculation order in the crossbar falls into two categories, horizontal and vertical. When the horizontal direction calculation mode is adopted, the OUs in the crossbar are calculated row by row, and the input WL vector may be reused. However, it is necessary to switch the ADC to quantify the MAC current of the same row of OUs.
When the vertical calculation mode is employed, there is no need to switch the ADCs, but it is necessary to update the input WL data in each calculation cycle. This design adopts a control signal $direction$ to decide which data flow mode is selected. 

As this work focuses on weight sparsity, input sparsity is not discussed in detail. Our proposed methodology can be applied alongside input sparsity techniques\cite{sre2019} without conflict. Once all the OUs have been computed, the new reordered data are then fed into Input Register of Computation Unit (IR-CU) for the next computation.  The longest computation time happens when there is no input sparsity detected, at that time the maximum number of computation cycles in each CU is the product of $Bitwidth_{input}$, $Crossbar_{height} // OU_{height}$ and $Crossbar_{width}//OU_{width}$.

\begin{figure}[H]
\centering
\includegraphics[width=8cm]{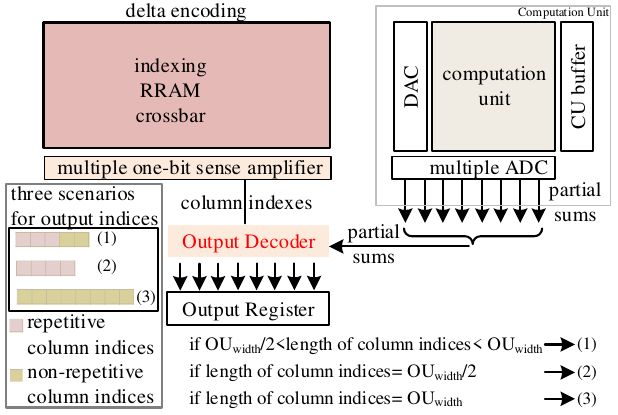}
\vspace{-0.2cm}
\caption{Output Routing Logic.}
\label{OUlogic}
\end{figure}

\textbf{Output Routing Logic: }Attention should be paid to the output indexing. We utilizes RRAM crossbars to store the binary formats of the column indices, as shown in Fig.~\ref{control} and Fig. \ref{OUlogic}. The important and only difference is that in this paper a single output value may need to be stored in two different output registers
% while in RePIM the one output data need to be stored only in one  output register. 
The repetitive columns are constrained to 2 in this paper. For an OU, the maximum number of column indices that need to be readout is 2$OU_{width}$, which occurs when all columns in the OU are repetitive columns. The minimum number of column indices that need to be readout is $OU_{width}$, which occurs when all columns in the OU are unique patterns. For OUs that contain both repetitive columns and non-repetitive columns, since the repetitive column indices are located before non-repetitive column indices (which is done by our algorithm), no additional resources are required to distinguish which indices are non-repetitive column indices and which are not. Thus the output decoder only needs to know the length of column indices to complete the output indexing. Meanwhile, to reduce the indexing storage overhead, delta encoding, which records the difference between column indices instead of their absolute values, is employed\cite{repim2021,effective2022}. The power overhead of the control logic (PE controller) shown in Fig. \ref{control} is only 0.48mW in a 32nm process, introducing only a slight overhead.
\vspace{-0.3cm}
\vspace{-0.4cm}

\vspace{-0.2cm}
\subsection{Comparison Results}
\begin{figure}[H]
\centering
\includegraphics[width=7.5cm]{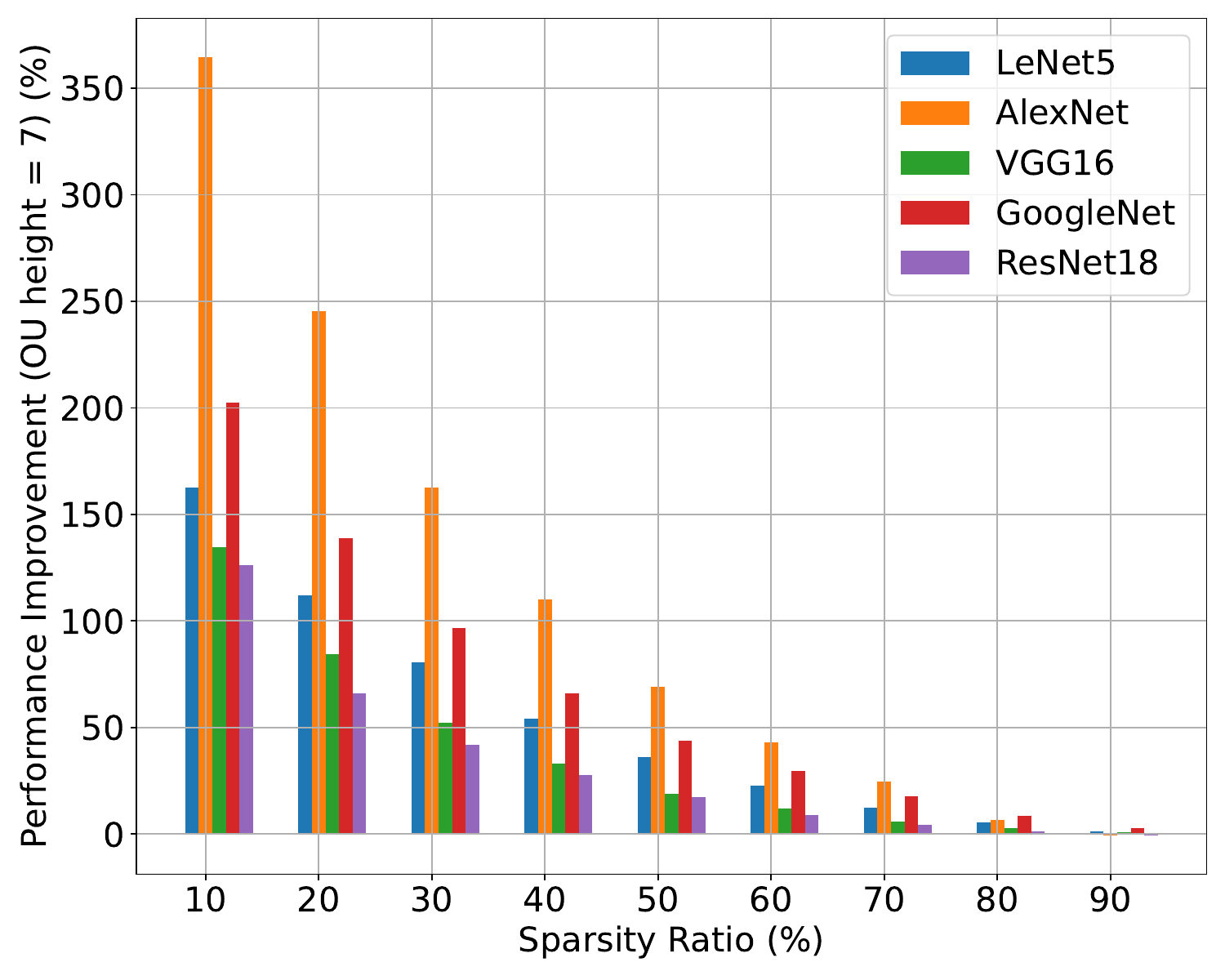}
\vspace{-0.3cm}
\caption{Performance improvement of five benchmarks with respect to RePIM\cite{repim2021}.}
\label{ou7modelwRes}
\end{figure}

Fig. \ref{ou7modelwRes} demonstrates the performance improvement achieved by the reordering scheme in this paper relative to the RePIM implementation. The performance is defined as the inverse of the product of computational crossbar quantities (CCQ) and energy consumption (EC), as shown in equation~\eqref{perf}.
\vspace{-0.2cm}
\begin{equation}
\label{perf}
\text{performance} = \frac{1}{\text{CCQ} \times \text{EC}}
\end{equation}
A reduction in computational crossbar resources or energy consumption directly enhances performance, as it reflects greater efficiency in completing the same inference tasks with fewer computational resources and lower power requirements. The hardware configuration details and power specifications of key components are shown in the Table \ref{hardware_detail}. Modifications occur only in the ADC resolution and OU size factoring in state-of-the-art readout circuits in RRAM\cite{tdc8b}\cite{repim2021}\cite{ADC40nm2022}. The average improvement of five typical models with respect to the RePIM with the value of $OU_{height}=7$ are \(54.15\%\) (LeNet5), \(113.92\%\) (AlexNet), \(38.27\%\) (VGG16), \(67.40\%\) (GoogleNet) and \(32.46\%\) (ResNet18) respectively, achieved an average of 61.24\% performance improvement.

Fig. \ref{ou7modelwRes} further reveals that our work's performance gain over RePIM becomes more pronounced at lower sparsity ratios. This occurs because at very high sparsity ratios (\(>\)80\%), gathering all-zero columns is relatively easy, where both approaches show comparable effectiveness with only marginal improvement from our design. However, at moderate-to-low sparsity, our solution delivers significant performance enhancements.
\vspace{-0.3cm}
\vspace{-0.4cm}

We further analyze the observed performance decrease at very high sparsity levels in comparison with RePIM. In our method, both all-zero and all-one rows are exploited simultaneously. For comparison, if we only consider all-zero rows, the probability of a single row being all zeros is
\vspace{-0.2cm}
\begin{equation}
P_{\mathrm{all-0}} = p^n.
\end{equation}\vspace{-0.1cm}
The probability of having at least $k$ all-zero rows among $m$ rows is given by \vspace{-0.3cm}
\begin{equation}\label{eq:cdf_all0}
P(X \ge k) = 1 - \sum_{i=0}^{k-1}  C_{m}^{i} (p^n)^i (1-p^n)^{\,m-i},
\end{equation}
where the random variable $X$ denotes the number of all-zero rows. When $p > 0.5$, zero bits dominate each column vector, so $p^n$ is significantly larger than $0.5^n$. Consequently, the expected number of all-zero rows is $\mathbb{E}[X] = m p^n$, which increases rapidly with $p$. If all-one rows are also included, the expected number of identical rows becomes $\mathbb{E}[X]=p^n + (1-p)^n$, which is slightly higher. However, when $p$ is very large ($p>0.5$), $(1-p)^n$ is negligible, so the difference between considering only all-zero rows and considering both all-zero and all-one rows is minor. This explains the observed decrease in performance benefits at very high sparsity, as illustrated in Fig.~\ref{ou7modelwRes}.

\begin{figure}[H]
\centering
\includegraphics[width=7.5cm]{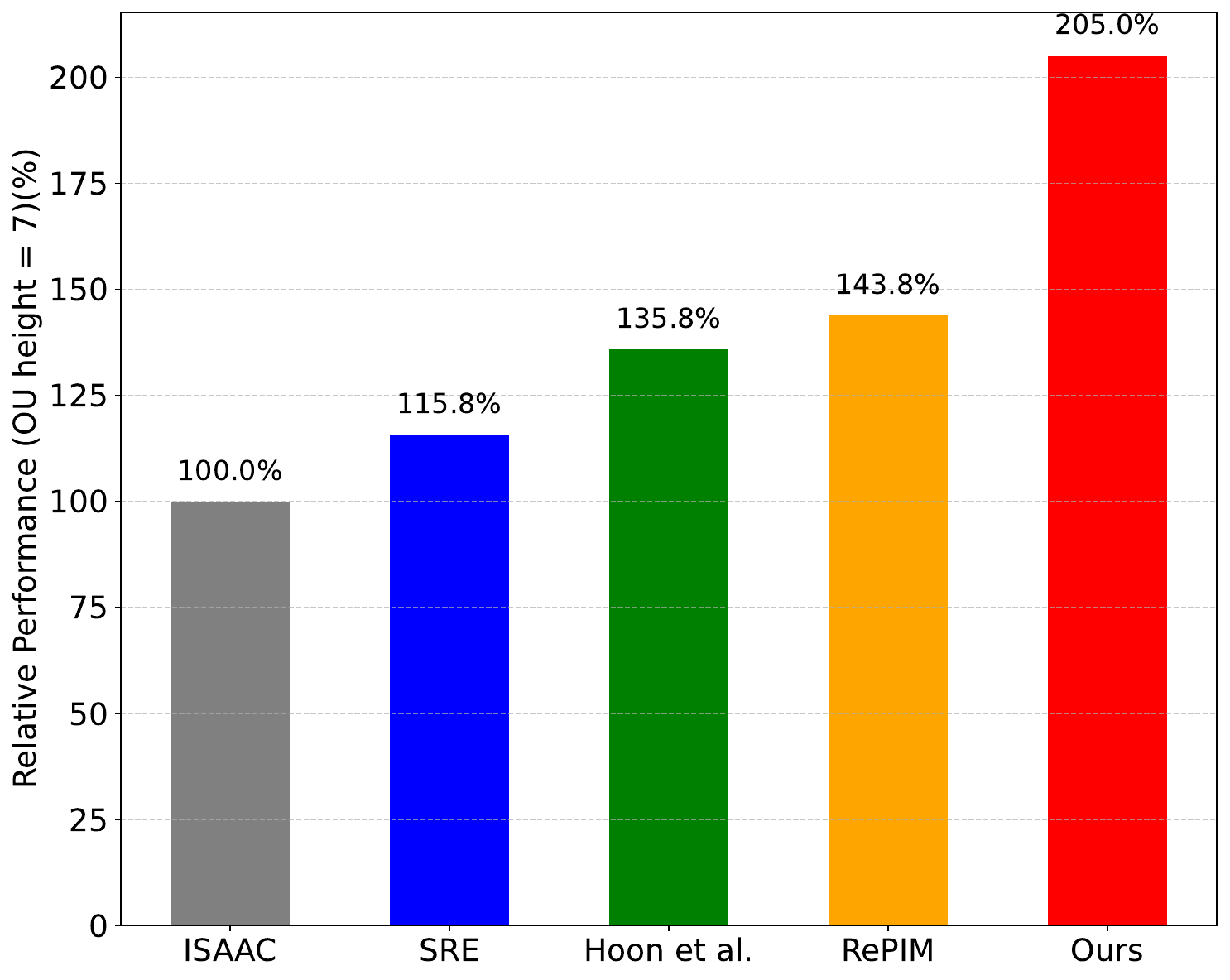}
\vspace{-0.2cm}
\caption{Performance improvement of five benchmarks with respect to the dense accelerator ISAAC\cite{isaac2016}.}
\label{ou7perform_comp}
\end{figure}

Comparison with other works can be seen in Fig. \ref{ou7perform_comp}. The over-idealized dense accelerator ISAAC\cite{isaac2016} is utilized as the baseline. The sparse accelerators, SRE\cite{sre2019}, RePIM\cite{repim2021} and Hoon et al.\cite{effective2022} are also evaluated as counterpart designs.

\begin{figure}[H]
\centering
\includegraphics[width=7.5cm]{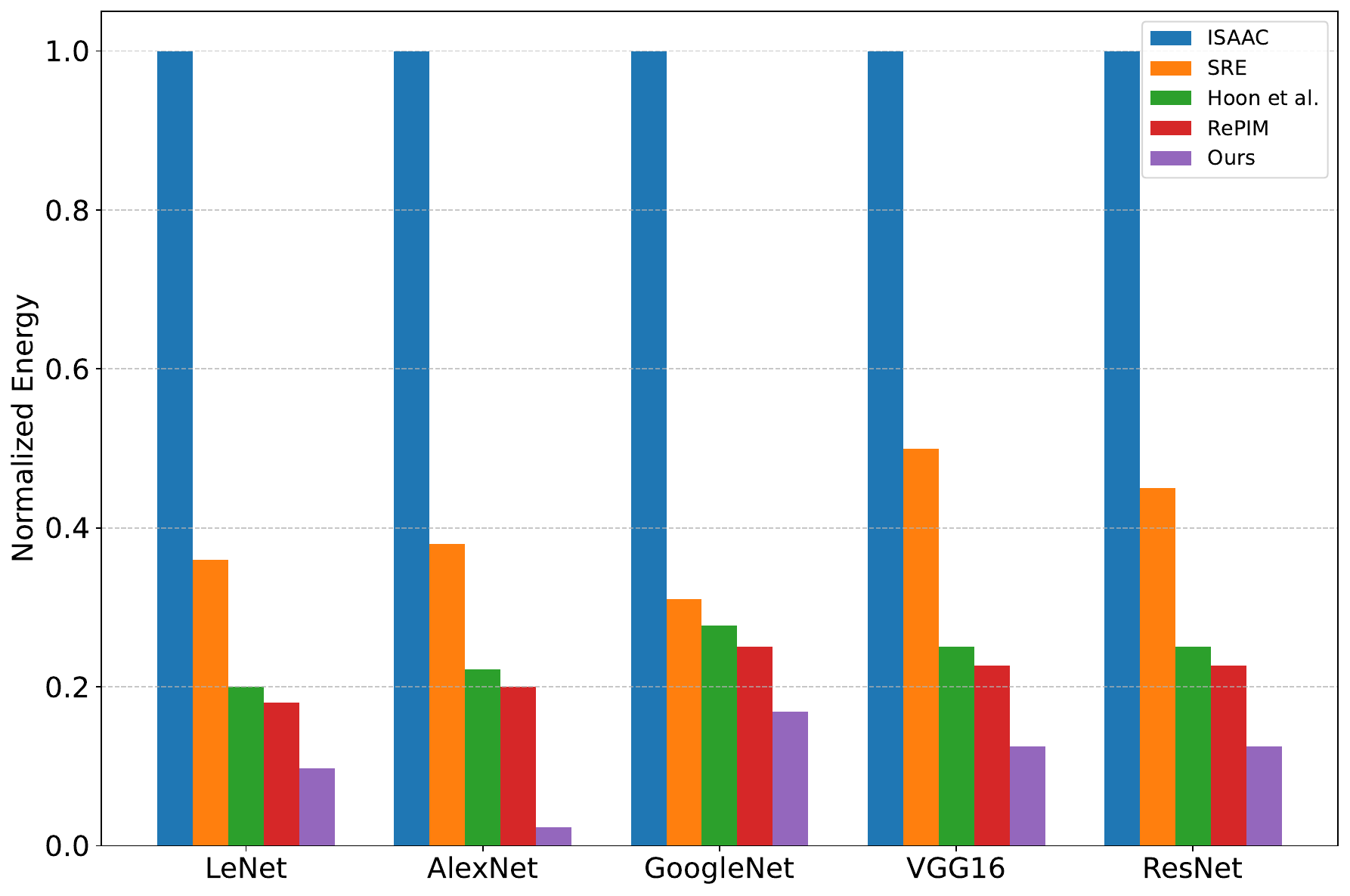}\vspace{-0.3cm}
\caption{Normalized energy consumption with respect to five benchmarks.}

\label{fig:energycomp}
\end{figure}

Regarding energy consumption, although the reordering algorithm introduces additional overheads from input reordering and output indexing, our design is still more energy-efficient than solutions without weight reordering. This advantage stems from two key factors. One is that our approach employs fewer crossbars compared to prior designs which requires similar indexing structures. This directly lowers the static energy consumption. Another is that the indexing operations on crossbars consume substantially less energy than computation-intensive operations\cite{repim2021}.  Although the output indexing in this work is more complex than in prior studies, the energy consumption it incurs is significantly smaller than the energy savings achieved through the proposed reordering strategy.  As a result, our design achieves a 1.51×–2.52× improvement in energy saving over RePIM. Based on the available data in  SRE\cite{sre2019}, RePIM\cite{repim2021} and Hoon et al.\cite{effective2022}, the normalized energy efficiency comparison is shown in Fig. \ref{fig:energycomp}, further demonstrating the superiority of our approach.
\vspace{-0.4cm}

\vspace{-0.7cm}
\vspace{-0.4cm}

It is important to note that our strategy introduces no accuracy loss in the neural networks, as it only repositions the weights within the crossbar without altering their values. This preserves model integrity and helps shorten time-to-market.

As for the comparison with digital CMOS-based bit-level optimization techniques, the energy consumption for the VGG-16 task is presented in Table \ref{latDigi}, where BitWave \cite{bitwave24} is used as the baseline. It is worth noting that the evaluation results take data movement overhead into account. Compared with CMOS-based designs, the energy efficiency improvement of our work benefits not only from the proposed bit-level sparsity optimization technique but also from the RRAM-based architecture, which significantly reduces the amount of data movement.
\vspace{-0.3cm}
\begin{table}[H]
\centering
\caption{Energy Comparison with CMOS-based works}
\vspace{-0.2cm}
\begin{tabular}{|c|c|c|c|c|c|}
\hline
        & \begin{tabular}[c]{@{}c@{}}Our\\ work\end{tabular} & \begin{tabular}[c]{@{}c@{}}BitCluster\\ \cite{bitcluster22}\end{tabular} & \begin{tabular}[c]{@{}c@{}}Bitlet\\ \cite{bitlet21, bitlet23t}\end{tabular} & \begin{tabular}[c]{@{}c@{}}BitWave\\ \cite{bitwave24}\end{tabular} & \begin{tabular}[c]{@{}c@{}}BBS\\ \cite{bbs24}\end{tabular} \\ \hline
\begin{tabular}[c]{@{}c@{}}Norm. Energy\\ Consumption\end{tabular}  & 0.5x& NaN& 1.02x& 1 & 0.62x                \\ \hline
\end{tabular}
\label{latDigi}
\end{table}

\vspace{-0.5cm}
\section{Conclusion}
\vspace{-0.1cm}
This paper proposes a bit level weight reordering strategy to support sparse NNs in RRAM-Acc. The strategy creates as many identical columns as possible through weight reordering and leverages the repetitiveness among these columns to save storage and reduce computations. Additionally, the two's complement format is adopted to replace the conventional positive-negative weight splitting method, thereby reducing crossbar overhead. A bit splitting policy is also employed to minimize output indexing overhead. Simulation results on five popular benchmarks demonstrate that our approach achieves a 61.24\% performance improvement compared to the state-of-the-art design.

\section*{Acknowledgement}
This work was supported in part by the National Natural Science Foundation of China under Grant U23A20322, Grant 61974164, Grant 62074166, Grant 62004219, Grant 62004220, Grant 62104256, and Grant 62304254.

\end{document}